\documentclass[12pt,preprint]{aastex}

\newcommand{\Msun}{\ensuremath{\mbox{M}_\odot}}

\newcommand{\gcc}{\ensuremath{\mathrm{g}~\mathrm{cm}^{-3}}}
\newcommand{\pcc}{\ensuremath{\mathrm{amu}~\mathrm{cm}^{-3}}}

\newcommand{\cHI}{\ensuremath{\mathrm{H}}}

\newcommand{\cHM}{\ensuremath{\mathrm{H}^{-}}}

\usepackage{natbib}
\usepackage{graphicx}

\begin{document}

\title{High-Entropy Polar Regions Around the First Protostars}
\author{Matthew J.~Turk\altaffilmark{1}, Michael L.~Norman\altaffilmark{1},
Tom Abel\altaffilmark{2}}
\altaffiltext{1}{Center for Astrophysics and Space Science, University of
California, San Diego}
\altaffiltext{2}{Kavli Institute for Particle Astrophysics and Cosmology,
Stanford University, 2575 Sand Hill Road, Menlo Park, CA 94025}

\begin{abstract}
We report on simulations of the formation of the first stars in the Universe,
where we identify regions of hot atomic gas ($f_{\mathrm{H}_2} < 10^{-6}$) at
densities above $10^{-14}~\gcc$, heated to temperatures ranging between 3000
and 8000 K.  Within this temperature range atomic hydrogen is unable to cool
effectively.  We describe the kinetic and thermal characteristics of these
regions and investigate their origin.  We find that these regions, while small
in total mass fraction of the cloud, may be dynamically important over the
accretion timescale for the central clump in the cloud, particularly as a
chemical, rather than radiative, mechanism for clearing the polar regions of
the accretion disk of material and terminating accretion along these
directions.  These inherently three-dimensional effects stress the need for
multi-dimensional calculations of protostellar accretion for reliable
predictions of the masses of the very first stars.
\end{abstract}
\maketitle


\section{Introduction}

As the physical model that governs simulations of metal-free stars expands to
include more chemical and radiative processes, the collapsing regions in which
they are expected to form have revealed a variety of interesting phenomena
\citep{2006ApJ...652....6Y,2008AIPC..990...16T,2010arXiv1006.1508C,2009Sci...325..601T,2010MNRAS.403...45S}.
The character of the inner regions of metal-free halos are strongly governed by
the reactions that govern the formation and dissociation of molecular hydrogen
at high densities \citep[submitted]{3Body}.  At densities below
$10^{-16}~\gcc$, the formation of molecular hydrogen is governed by
electron-catalyzed association of \cHM and \cHI
\citep{abel97,2008MNRAS.388.1627G}.   However, at densities of $10^{-16}~\gcc$
and above, the formation of molecular hydrogen is dominated by three-body
reactions, primarily where the third body is atomic hydrogen:
$$\begin{array}{lcl}
\mathrm{H} + 2\mathrm{H} & \rightarrow & \mathrm{H} + \mathrm{H}_2 \\
\mathrm{H} + \mathrm{H}_2 & \rightarrow & \mathrm{H} + 2\mathrm{H}.
\end{array}$$
The reaction rates for these two reactions have been shown not only to be
uncertain to an order of magnitude or more \citep{2008AIPC..990...25G} but
these uncertainties may have a substantial impact on the structure and
mass-distributions of collapsing primordial clouds \citep{3Body}.  Every
molecule formed through this reaction releases $4.48~\mathrm{eV}$ in thermal
energy into the gas, and because the rate of formation decreases with
increasing temperature, the overall process of transforming the atomic hydrogen
gas in into a fully-molecular state occurs over several decades in
density \citep{RA04,1998ApJ...508..141O,1998A&A...335..403G,PSS83,2008AIPC..990...25G}.
In 1D, spherically symmetric calculations it is inevitable that all
high-density gas undergoes this transition.  However, in full 3D calculations,
the mechanism of transformation is more complex, owing to variations in the
density-temperature distribution and the velocity structure of the collapsing
cloud.

We present results of high-resolution computer simulations of metal-free star
forming regions.  In these simulations we identify regions of high-density
($>10^{-15}~\gcc$) atomic gas whose presence may contribute to a chemical,
rather than radiative, mechanism for slowing or terminating accretion along the
poles of primordial accretion disks \citep{2008ApJ...681..771M}.

\section{Simulations}\label{sec:simulations}

We report on a single simulation drawn from a suite of calculations, designed
to sample a variety of formation environments.  We centered the simulation
volume on the location of the peak density of the first dark matter halo of
mass $10^6~\Msun$ to form and generated three nested subgrid levels, for an
initial effective resolution of $1024^3$.  Our initial conditions are generated
as in \cite{ABN02,oshea07a}, with the notable exception that we have updated
our cosmological parameters to WMAP 7 \citep{2010arXiv1001.4744J, PTMain}.
These simulations were conducted using the code Enzo
\citep{oshea04,1998ApJ...495...80B,2001astro.ph.12089B}, using a 9-species
primordial chemistry model implementing the \texttt{TWOSTEP} method described
in \cite{Verwer94gauss-seideliteration} and in the appendix to
\cite{2009Sci...325..601T}.  While we focus on the effects of three-body
molecular hydrogen formation with atomic hydrogen acting as the third body, we
also solve the set of rate equations where molecular hydrogen acts as the third
body.  At densities of approximately $10^{-15}-10^{-13}~\gcc$, primordial gas
becomes optically thick to the cooling of molecular hydrogen from
ro-vibrational transitions.  In our simulations, this is applied using the
prescription given in \cite{RA04}, which agrees well with the detailed Sobolev
approximation \citep{3Body,sobolev1960,2006ApJ...652....6Y}.  The cooling
radiation is reduced by:
$$
L_{\mathrm{lines,thick}}(T) =
L_{\mathrm{lines,thin}}\times\mathrm{min}(1,(n/n_0)^{-\beta}),
$$
with $n_0 = 8\times10^{9}~\pcc$ and $\beta = -0.45$ \citep{RA04}.

We discuss a relatively slow-collapsing halo that we can examine with some
depth, but this halo is not extraordinary, nor does it fragment
\citep{2009Sci...325..601T,2010MNRAS.403...45S,2010arXiv1006.1508C}.
Furthermore, the phenomena discussed in this paper are present in a suite of
simulations, and their frequency will be discussed in the forthcoming
\cite{PTMain}.  The final output of the calculation at $z = 17.4$, has peak
density of $2.0\times10^{-9}~\gcc$ at the 29th level of refinement.  The entire
simulation has $3.3\times10^7$~computational elements, including
$2.3\times10^{6}$ with cell size less than $35~\mathrm{AU}$.

We also introduce an improved refinement criterion as compared to previous
simulations. We require the numerical resolution to be one sixteenth of the
local Jeans length. However, the calculation of the Jeans length is done
assuming the gas is at $T_{min} = 200~\mathrm{K}$ even when it has not cooled
so low. As a consequence we have $(T/200{\rm K})^{9/2}$ more resolution
elements in the material before it forms the cold phase. We refer to this as
the cold Jeans refinement criterion, which for all the material above the
minimum temperature is more stringent then the standard criterion. This
modification ensures high resolution at high densities even before the material
forms the cooling cores.  In addition, we also refine on overdensity in both
baryonic and dark matter content at a factor of 4.0 which ensures high
resolution in the lower density material that forms first star hosting mini
halos.

\section{Results}\label{sec:results}

\begin{figure*}
\begin{centering}
\includegraphics[width=0.95\textwidth]{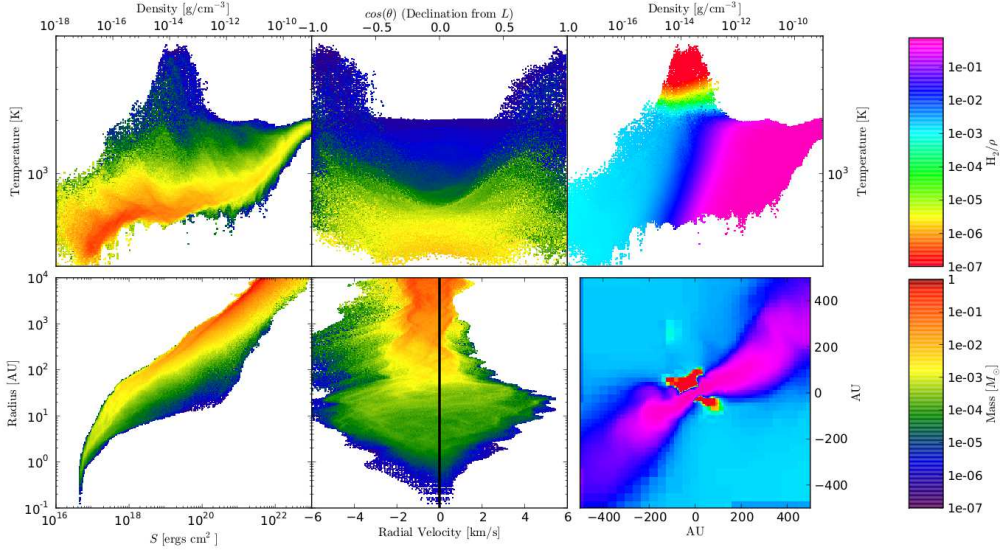}
\caption{We show the distributions of the thermodynamic and kinetic states of
the inner $10^{4}~\mathrm{AU}$ of a collapsing primordial Population III
star-forming region gas as a function of different variables:  mass
distribution as a function of density and temperature (upper left); mass
distribution as a function of the declination (co-latitude) angle from the
local angular momentum vector and temperature (upper middle); mass-weighted
average molecular hydrogen mass fraction as a function of density and
temperature (upper right); mass distribution as a function of radius from the
center and an entropy-like quantity (lower left); mass distribution as a
function of the radius and the local radial velocity of the gas (lower middle).
In the lower right panel, we show a slice through the center of the simulation
of the molecular hydrogen mass fraction.  The field of view in the slice image
is $1000~\mathrm{AU}$.  Colorbars showing the molecular hydrogen fraction in
each of the molecular hydrogen fraction plots and the mass in each pixel of the
mass-distribution plots have been placed on the far right, in the upper and
lower panels respectively.}
\label{fig:unified}
\end{centering}
\end{figure*}

\begin{figure*}
\begin{centering}
\includegraphics[width=0.95\textwidth]{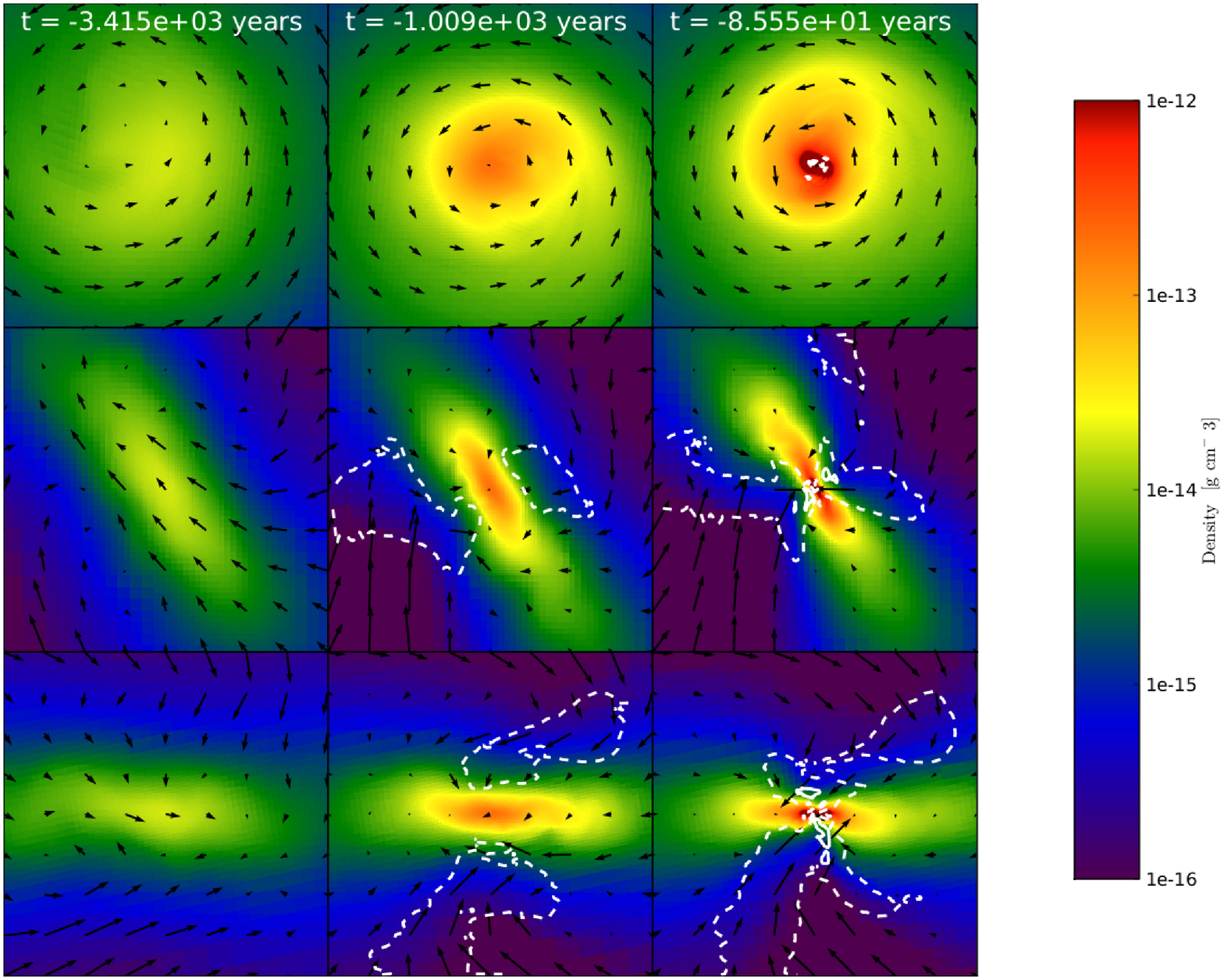}
\caption{Slices at three different epochs (columns) in the collapse of the
primordial cloud.  The field of view in all images is $2000~\mathrm{AU}$ on a
side.  The axes of alignment of the three rows have been chosen in an
orthogonal frame (not the computational axes) where the normal
vector of the top image is aligned with the angular momentum vector of the
innermost $2000~\mathrm{AU}$ of the disk, and the middle and lower rows are
mutually orthogonal.  Overplotted are velocity vectors and a contour at
$T=1000~\mathrm{K}$.  From left, the times of the images are 3415, 1700, 1000,
320, and 85 years prior to the final output of the simulation.}
\label{fig:cpimages}
\end{centering}
\end{figure*}

As the gas collapses and is converted to molecular hydrogen, we see a spread in
the temperature and the molecular hydrogen fraction as a function of density,
as shown in the upper left and upper right panels of Figure~\ref{fig:unified}.
At densities between approximately $10^{-15}~\gcc$ and $10^{-10}~\gcc$ the mass
is distributed over a broad range of temperatures, while the corresponding
spread in molecular hydrogen fractions is an order of magnitude or more.  This
results in a broad variation of the rate coefficients governing the transition
to molecular hydrogen.  For higher temperature gas, the overall transition
occurs more slowly, resulting in less efficient ro-vibrational cooling.  In
some regions of phase space, the onset of the transformation of the gas is
delayed until the onset of optical thickness.

In the upper left panel of Figure~\ref{fig:unified} we plot the distribution of
mass as a function of density and temperature.  The clear trend is for the gas
to follow standard tracks through phase space, similar to those shown in
\cite{ABN02,2006ApJ...652....6Y,2009Sci...325..601T,oshea07a}.  However, we
note the strong presence of low-mass regions of high-temperature gas,
particularly at densities between $10^{-15}~\gcc$ and $10^{-13}~\gcc$.

In the upper middle panel of Figure~\ref{fig:unified} we have plotted the
distribution of mass in the innermost $10^{4}~\mathrm{AU}$ of the halo as a
function of the temperature and the cosine of declination from the average
angular momentum vector at that radius.  We have chosen cosine to ensure equal
latitudinal area across the entire domain.  In the normalization used here, the
two extremes ($-1.0$ and $1.0$) correspond to the polar regions and the middle
region ($0.0$) corresponds to the disk region.  All gas above a temperature of
roughly $2000~\mathrm{K}$ is confined within 60 degrees from the poles defined
by the local angular momentum vector.

The upper right panel of Figure~\ref{fig:unified} shows the mass-weighted
average molecular hydrogen mass fraction as a function of density and
temperature.  All of the gas above $2000~\mathrm{K}$ is mostly or completely
dissociated and free of molecules.  The lower-temperature gas follows a typical
transition from atomic to molecular states.

In the lower left panel of Figure~\ref{fig:unified} we have plotted the
distribution of mass with respect to radius from the densest point of our cloud
(x-axis) and an entropy-like quantity ($S$, y-axis) defined by

$$S = \frac{k_{\mathrm{b}}}{m_\mathrm{H}} \frac{T}{\mu\rho^{2/3}},$$

where $T$ is the temperature, $\mu$ is the mean molecular weight and $\rho$ is
the density.  Between $10~\mathrm{AU}$ and $100~\mathrm{AU}$ the
high-temperature regions are visible as a broadening in the entropy profile,
indicating a difference in the entropy gradient within the collapsing region.

In the lower middle panel of Figure~\ref{fig:unified} we plot the mass
distribution as a function of distance from the densest zone and the radial
velocity with respect to that zone.  The velocity of the central zone in the
calculation has been subtracted prior to calculation of the radial velocity.
Several independent flow lines are visible, including several with net outward
motion, particularly within $\sim50~\mathrm{AU}$.  The bulk flow is inward and
the spread in velocities very close to the central point is expected as a
result of both rotation and incoherent flow.

The lower right panel of Figure~\ref{fig:unified} gives a slice through the
core, aligned with the y-axis, $1000~\mathrm{AU}$ on a side, where the color
corresponds to the molecular hydrogen mass fraction.  We note the obvious
molecular hydrogen-dominated disk and atomic high temperature polar regions.

In Figure~\ref{fig:cpimages}, we display slices through the primordial cloud
that are aligned with a new orthogonal reference frame such that the normal to
the image plane is aligned with the angular momentum vector of the innermost
$2000~\mathrm{AU}$ of the region.  Overlaid are velocity vectors and contours
at temperatures of $1000~\mathrm{K}$ (dotted) and $2000~\mathrm{K}$ (solid).
Each column corresponds to a different epoch in the cloud's collapse, where
time of $0.0~\mathrm{years}$ is the final output of the calculation.  The upper
panel shows disk settling  at densities of approximately $10^{-14}~\gcc$.  This
disk undergoes little evolution over the displayed epochs in the top, face-on
view, In the middle and lower panels, gas is flowing along the poles of the
disk to the central point.  High-temperature gas develops in these polar
regions, at and above $2000~\mathrm{K}$, although we note that the largest
pockets of hot gas are at $1000~\mathrm{K}$ and extend nearly
$1000~\mathrm{AU}$.

\section{Discussion}\label{sec:discussion}

\subsection{Chemical Origin and Stability}

The formation of molecular hydrogen at densities greater than $10^{-16}~\gcc$
is dominated by the three-body formation mechanism, where two hydrogen atoms
and a catalyst produce a single molecule of hydrogen.  Formation through this
channel produces an excited molecule of hydrogen, which rapidly collisionally
de-excites to its ground state, depositing $4.48~\mathrm{eV}$ of thermal energy
in the gas.  Conversely, collisional dissociation of a molecule of hydrogen
removes from the gas $4.48~\mathrm{eV}$ of thermal energy.  As the formation
and destruction rates of molecular hydrogen via these processes decrease and
increase, respectively, with the temperature of the gas, this reduction in
thermal energy carries with it a corresponding reduction in the rate at which
molecular hydrogen associates.  This process can be viewed as an increase in
the specific heat of the gas; every molecule of hydrogen that is to be
dissociated requires the introduction of $4.48~\mathrm{eV}$ of energy.  The
equilibrium density of atomic hydrogen, at a fixed temperature and subject only
to the three-body process and its inverse, can be written as:
\begin{equation}
n_\mathrm{H,eq} = \frac{-1}{4 k_{22}}(k_{13} - \sqrt{8k_{13}k_{22}
n_{\mathrm{H}, \mathrm{H_2}} + k_{13}^2}).\label{eq:equilibrium}
\end{equation}
where $k_{13}$ is the dissociation rate in units of
$\mathrm{cm}^3~\mathrm{s}^{-1}$ and $k_{22}$ is the association rate in
$\mathrm{cm}^3~\mathrm{s}^{-1}$, $n_\mathrm{H,eq}$ is the atomic hydrogen density
in $cm^{-3}$ and $n_{\mathrm{H}, \mathrm{H_2}}$ is the number density of the
mixed hydrogen gas.  In our simulation, we used values for $k_{22}$ from
\citet{2008AIPC..990...25G} and $k_{13}$ from \citet{MSM96}.  As shown in the
upper right panel of Figure~\ref{fig:unified}, when gas exceeds roughly
$2300~\mathrm{K}$ at a density of $10^{-14}~\gcc$, the destruction rate
dominates and the gas rapidly dissociates, but reaching or breaching that
temperature barrier requires an influx of energy to dissociate any molecules of
hydrogen.  We note that while the temperature at which dissociation dominates
will be weakly dependent on density, as seen in Equation~\ref{eq:equilibrium}.
Having less energy ``locked up'' in hydrogen molecules lowers the energy
barrier to dissociation, as the gas can shed less energy via dissociative
cooling.

In the absence of shocks, asymmetries or other perturbations of the gas, as in
spherically-symmetric, 1D calculations, molecular hydrogen that is allowed to
cool will simply slow its collapse if its temperature reaches the temperature
at which the dissociative process dominates.  The gas will loiter until
radiative processes are able to shed enough thermal energy to allow it to
collapse further.  However, in our simulations, the gas may be subject to
asymmetric inflow, minor shocks and disk formation, as discussed in
\cite{ABN02, 2009Sci...325..601T, 2010arXiv1006.1508C, 2010MNRAS.403...45S,
oshea07a}.  Additionally, our physical model includes optical thickness to
ro-vibrational cooling and our simulation resolves inhomogeneities of the
density and temperature of the gas, both factors that change this evolution.

With a smaller specific heat as a result of a lower molecular fraction, the
molecular hydrogen can be dissociated by minor shocks and other perturbations
from infalling gas; once it has reached high temperatures, the energy barrier
to associating molecular hydrogen is just as high, but with no mechanism for
shedding thermal energy until compressional heating drives the temperature to
$\sim10^{4}~\mathrm{K}$.  In the time-series plots in Figure~\ref{fig:cpimages}
we show that the flow, while largely quiescent, features strong asymmetric
inflow as well as fast infall along the poles.  This produces a shocked,
high-temperature gas which is subsequently compressionally heated.

At this point, the gas will be able to cool efficiently enough that it will
likely not heat further, but it will also be unable to form substantial amounts
of molecular hydrogen.  This transition is shown in the upper right panel of
Figure~\ref{fig:unified}, where the high-temperature, shocked regions in the
upper left panel are shown to have fully dissociated their reservoir of
molecular hydrogen, and thus the only available coolant.  We also draw
attention to the power-law stratification in the molecular hydrogen fraction as
a function of density and temperature, indicating an equilibrium molecular
hydrogen fraction at those temperatures.

In the bottom left panel of Figure~\ref{fig:unified}, we can identify two
features in the entropy of the gas.  The first is that the entropy overall
decreases with decreasing radii; the cloud is globally stable against
convection, which would require a reversal of the entropy gradient.  However,
the warm gas behaves adiabatically, retaining essentially constant entropy with
decreasing radius.  During subsequent accretion onto the protostellar cloud,
this gas will continue to be confined by colder, radiatively efficient gas.
Although globally stable against convection, at later times these adiabatic
regions may undergo convective instability.  In the final output plotted in
Figure~\ref{fig:cpimages} (bottom rightmost image) we note that a pocket of gas
of $\sim1000~\mathrm{K}$ has become distended and nearly disconnected from its
innermost component; this may be the start of a buoyant path, up and to the
right in the image plane.

\subsection{Radial Velocity and Disk Structure}

In the lower middle panel of Figure~\ref{fig:unified}, we can identify several
distinct ``tracks'' along which gas is flowing.  Each of the hot, dissociated
regions is characterized by amplification of the inward radial velocity and a
relatively small radial extent.  Gas is falling inward onto the hot atomic
bubbles, where it dissociates, increases in pressure, and then builds up on the
inward side of the velocity spikes.  This variance in the radial velocity of
the shocked, dissociated regions results in gas that is collapsing less slowly;
in the comoving frame of its surroundings, it may even be rising, although we
cannot determine this at the stopping point in our simulation.

As seen in Figure~\ref{fig:cpimages}, in this simulation a clear disk is
formed, accentuated by the strong differences in the chemical and thermal
states of the gas in the disk and in the polar regions.  Examining the angular
distribution of gas (Figure~\ref{fig:unified}, upper middle panel) we see that
the shocks are essentially confined to within or near the polar regions of the
disk, demonstrated as well in the lower right panel of
Figure~\ref{fig:unified}.  The mechanism by which Population III protostellar
feedback will break out of the disk, as well as the density structure of a
settled disk both require an understanding of this material, as the gas flowing
along the poles will necessarily interact with these high-entropy regions.  In
past theoretical works discussing the accretion mechanism onto primordial
protostars, radiative breakout has been identified as the sole mechanism by
which accreting material flowing along the poles could be slowed or stopped
\citep{2008ApJ...681..771M}.  This scenario may change if the polar regions are
largely composed of high-temperature, atomic gas that is buoyant or unstable to
convection.  We propose that a chemical component may contribute to changes in
the accretion flow even before the onset of radiative feedback from the central
protostar, at times later than those probed in this simulation.  The decreased
molecular opacities in the polar region will allow protostellar accretion
luminosity to reach larger radii, feeding back on the accretion flow.

\section{Conclusions}\label{sec:conclusions}

The hydrodynamics of disk formation around a protostar lead to high temperature
atomic polar regions.  Until compressional heating drives the temperature of
these hot, dissociated regions to $10^4~\mathrm{K}$, they have no opportunity
for radiative cooling.  The initial formation of these bubbles is through
shocks dominated by large scale flows, and the chemical processes that govern
their evolution will continue to enable their formation.  We see these
dissociated regions prior to the formation of the central protostar, and we
extrapolate that they will continue to persist until the protostar begins to
accrete.  Clearing material from the polar regions of the accretion disk,
previously thought to be exclusively through radiative feedback mechanisms, is
necessary to reverse accretion flows along the polar directions and enable the
breakout of ionizing radiation.  However, the process of terminating accretion
along the polar regions of the accretion disk may be assisted by these hot
regions, either by the dredging of material via buoyant or convective rising of
material, or simply through the chemical and kinetic processes that prevent the
region from associating molecular hydrogen.  This may contribute to a much
earlier breakout of radiation from the protostellar core than found in the
models of \citet{2008ApJ...681..771M}, although (as noted in that paper) the
total mass of gas flowing along the poles is much lower than that accreted from
the disk.

The spread in the chemical, thermal and kinetic states of the gas as it regimes
such as where we see these shocking regions, must be resolved to understand the
mechanism by which accretion disks develop and matter is processed.
One-dimensional codes, or codes that assume a universal equation of state as a
function of density, are unable to do so.  Furthermore, simulation codes that
are unable to resolve the development and persistence of shocks, or that are
otherwise unable to apply adequate resolution, are equally unable to resolve
this formation and settling of a disk and its subsequent evolution.  Future
simulations that bypass the Courant condition, such as sink particles (as in
\cite{2010arXiv1006.1508C, 2010MNRAS.403...45S}) while still resolving the
relevant chemical and hydrodynamical processes, will be necessary to study the
long term impact of these hot, atomic polar regions.

\acknowledgments M.J.T.\ acknowledges support by NASA ATFP grant NNX08AH26G
and NSF AST-0807312.  These simulations were conducted utilizing the Triton
Resource at San Diego Supercomputer Center.  The authors thank Greg Bryan,
Jeff Oishi, Britton Smith and Brian O'Shea for helpful comments, as well as
Stephen Lepp, David Collins, Paul Clark, Simon Glover and Ralf Klessen for
productive discussions.  We thank the anonymous referee for thoughtful comments
and suggestions.


\begin{thebibliography}{24}
\expandafter\ifx\csname natexlab\endcsname\relax\def\natexlab#1{#1}\fi

\bibitem[{{Abel} {et~al.}(1997){Abel}, {Anninos}, {Zhang}, \&
  {Norman}}]{abel97}
{Abel}, T., {Anninos}, P., {Zhang}, Y., \& {Norman}, M.~L. 1997, New Astronomy,
  2, 181

\bibitem[{{Abel} {et~al.}(2002){Abel}, {Bryan}, \& {Norman}}]{ABN02}
{Abel}, T., {Bryan}, G.~L., \& {Norman}, M.~L. 2002, Science, 295, 93

\bibitem[{{Bryan} {et~al.}(2001){Bryan}, {Abel}, \&
  {Norman}}]{2001astro.ph.12089B}
{Bryan}, G.~L., {Abel}, T., \& {Norman}, M.~L. 2001, ArXiv Astrophysics
  e-prints

\bibitem[{{Bryan} \& {Norman}(1998)}]{1998ApJ...495...80B}
{Bryan}, G.~L. \& {Norman}, M.~L. 1998, \apj, 495, 80

\bibitem[{{Clark} {et~al.}(2010){Clark}, {Glover}, {Klessen}, \&
  {Bromm}}]{2010arXiv1006.1508C}
{Clark}, P.~C., {Glover}, S.~C.~O., {Klessen}, R.~S., \& {Bromm}, V. 2010,
  ArXiv e-prints

\bibitem[{{Galli} \& {Palla}(1998)}]{1998A&A...335..403G}
{Galli}, D. \& {Palla}, F. 1998, \aap, 335, 403

\bibitem[{{Glover}(2008)}]{2008AIPC..990...25G}
{Glover}, S. 2008, in American Institute of Physics Conference Series, Vol.
  990, First Stars III, ed. B.~W. {O'Shea}, A.~{Heger}, \& T.~{Abel}, 25--29

\bibitem[{{Glover} \& {Abel}(2008)}]{2008MNRAS.388.1627G}
{Glover}, S.~C.~O. \& {Abel}, T. 2008, \mnras, 388, 1627

\bibitem[{{Jarosik} {et~al.}(2010){Jarosik}, {Bennett}, {Dunkley}, {Gold},
  {Greason}, {Halpern}, {Hill}, {Hinshaw}, {Kogut}, {Komatsu}, {Larson},
  {Limon}, {Meyer}, {Nolta}, {Odegard}, {Page}, {Smith}, {Spergel}, {Tucker},
  {Weiland}, {Wollack}, \& {Wright}}]{2010arXiv1001.4744J}
{Jarosik}, N., {Bennett}, C.~L., {Dunkley}, J., {Gold}, B., {Greason}, M.~R.,
  {Halpern}, M., {Hill}, R.~S., {Hinshaw}, G., {Kogut}, A., {Komatsu}, E.,
  {Larson}, D., {Limon}, M., {Meyer}, S.~S., {Nolta}, M.~R., {Odegard}, N.,
  {Page}, L., {Smith}, K.~M., {Spergel}, D.~N., {Tucker}, G.~S., {Weiland},
  J.~L., {Wollack}, E., \& {Wright}, E.~L. 2010, ArXiv e-prints

\bibitem[{{Martin} {et~al.}(1996){Martin}, {Schwarz}, \& {Mandy}}]{MSM96}
{Martin}, P.~G., {Schwarz}, D.~H., \& {Mandy}, M.~E. 1996, \apj, 461, 265

\bibitem[{{McKee} \& {Tan}(2008)}]{2008ApJ...681..771M}
{McKee}, C.~F. \& {Tan}, J.~C. 2008, \apj, 681, 771

\bibitem[{{Omukai} \& {Nishi}(1998)}]{1998ApJ...508..141O}
{Omukai}, K. \& {Nishi}, R. 1998, \apj, 508, 141

\bibitem[{{O'Shea} {et~al.}(2004){O'Shea}, {Bryan}, {Bordner}, {Norman},
  {Abel}, {Harkness}, \& {Kritsuk}}]{oshea04}
{O'Shea}, B.~W., {Bryan}, G., {Bordner}, J., {Norman}, M.~L., {Abel}, T.,
  {Harkness}, R., \& {Kritsuk}, A. 2004, ArXiv Astrophysics e-prints

\bibitem[{{O'Shea} \& {Norman}(2007)}]{oshea07a}
{O'Shea}, B.~W. \& {Norman}, M.~L. 2007, \apj, 654, 66

\bibitem[{{Palla} {et~al.}(1983){Palla}, {Salpeter}, \& {Stahler}}]{PSS83}
{Palla}, F., {Salpeter}, E.~E., \& {Stahler}, S.~W. 1983, \apj, 271, 632

\bibitem[{{Ripamonti} \& {Abel}(2004)}]{RA04}
{Ripamonti}, E. \& {Abel}, T. 2004, \mnras, 348, 1019

\bibitem[{{Sobolev}(1960)}]{sobolev1960}
{Sobolev}, V.~V. 1960, {Moving envelopes of stars} (Cambridge: Harvard
  University Press, 1960)

\bibitem[{{Stacy} {et~al.}(2010){Stacy}, {Greif}, \&
  {Bromm}}]{2010MNRAS.403...45S}
{Stacy}, A., {Greif}, T.~H., \& {Bromm}, V. 2010, \mnras, 403, 45

\bibitem[{{Turk} {et~al.}(2010{\natexlab{a}}){Turk}, {Abel}, \&
  {Norman}}]{PTMain}
{Turk}, M.~J., {Abel}, T., \& {Norman}, M.~L. 2010{\natexlab{a}}, \apj, in
  Preparation

\bibitem[{{Turk} {et~al.}(2009){Turk}, {Abel}, \&
  {O'Shea}}]{2009Sci...325..601T}
{Turk}, M.~J., {Abel}, T., \& {O'Shea}, B. 2009, Science, 325, 601

\bibitem[{{Turk} {et~al.}(2008){Turk}, {Abel}, \&
  {O'Shea}}]{2008AIPC..990...16T}
{Turk}, M.~J., {Abel}, T., \& {O'Shea}, B.~W. 2008, in American Institute of
  Physics Conference Series, Vol. 990, First Stars III, ed. B.~W. {O'Shea},
  A.~{Heger}, \& T.~{Abel}, 16--20

\bibitem[{{Turk} {et~al.}(2010{\natexlab{b}}){Turk}, {Clark}, {Glover},
  {Greif}, {Abel}, {Klessen}, \& {Bromm}}]{3Body}
{Turk}, M.~J., {Clark}, P.~C., {Glover}, S.~C.~O., {Greif}, T.~H., {Abel}, T.,
  {Klessen}, R.~S., \& {Bromm}, V. 2010{\natexlab{b}}, \apj, submitted (June,
  2010)

\bibitem[{Verwer(1994)}]{Verwer94gauss-seideliteration}
Verwer, J.~G. 1994, SIAM J. Sci. Comput, 15, 1243

\bibitem[{{Yoshida} {et~al.}(2006){Yoshida}, {Omukai}, {Hernquist}, \&
  {Abel}}]{2006ApJ...652....6Y}
{Yoshida}, N., {Omukai}, K., {Hernquist}, L., \& {Abel}, T. 2006, \apj, 652, 6

\end{thebibliography}
\end{document}